\documentclass{article}
\usepackage{spconf,amsmath,graphicx,hyperref,booktabs, bbding, pifont}
\usepackage{balance, multirow}
\usepackage{tabularx}
\usepackage{amssymb}
\usepackage[table]{xcolor}
\usepackage{makecell}
\hypersetup{
    colorlinks=false,
    hidelinks
}

\title{Rethinking Post-Training for Expressive Zero-Shot Text-to-Speech}

\name{\shortstack{Jingbin Hu$^{1}$, Xinfa Zhu$^{1}$, Hao Wang$^{1}$, Wenjie Tian$^{1}$, Kangxiang Xia$^{1}$, Hangrui Hu$^{1}$, \\ Kexin Huang$^{1}$, Jin Xu$^{1,\dag}$, Liumeng Xue$^{2,3,\dag}$, Lei Xie$^{1,\dag}$}\thanks{$^\dag$ Corresponding author.}}
\address{
$^1$ Alibaba Token Hub, Alibaba Group \\
$^2$ State Key Laboratory of Novel Software Technology, Nanjing University, Nanjing, China \\
$^3$ School of Intelligence Science and Technology, Nanjing University, Suzhou, China
}

\begin{document}
\ninept
\maketitle
\begin{abstract}
Zero-shot voice cloning typically emphasizes preserving the timbre and prosodic characteristics of a reference speaker. However, in real-world use, speech should adapt its delivery to the target text rather than remain tied to a fixed speaking style. In this work, we formulate this problem as text-adaptive prosody synthesis for zero-shot voice cloning and study how post-training can endow a cloning model with this capability. We develop a three-stage framework. First, Direct Preference Optimization (DPO) provides a cold start by aligning the model with text-appropriate expressive speech. Second, Group Sequence Policy Optimization (GSPO) jointly optimizes prosodic expressiveness, acoustic quality, intelligibility, and speaker similarity. Finally, context-conditioned On-Policy Self-Distillation (OPSD) enables self-improvement by transferring expressive behavior from strong-reference to weak-reference conditions. On Qwen3-TTS, our framework improves Emotion from 18.5 to 81.2 and Paralinguistics from 38.4 to 73.9. Extensive ablations validate the role of each stage. Speech samples are available~\footnote{Demo:\href{https://hujingbin1.github.io/TTS-Post-Train/}{https://hujingbin1.github.io/TTS-Post-Train/}}.
\end{abstract}


\section{INTRODUCTION}

Recent advances in zero-shot text-to-speech (TTS) enable speech generation from a target text and a short reference utterance, making voice cloning increasingly practical~\cite{cosyvoice3,moss-tts,qwentts,voxcpm,f5-tts,indextts}. A central goal of zero-shot voice cloning is to preserve the characteristics of the reference speaker, including speaker timbre and speaking style. However, faithfully inheriting the prosodic characteristics of a reference is not always desirable. In real-world scenarios, the reference utterance primarily serves as evidence of who is speaking, while the target text determines how the speech should be delivered. A reference with limited intonational variation, for example, may lead to flat speech even when the target text calls for strong emotion, emphasis, or paralinguistic expression. This raises a fundamental question: \textbf{can a voice cloning model preserve the speaker identity while adapting its prosody to the target text, rather than simply imitating the reference speaking style?}

We study this problem as \textbf{text-adaptive prosody synthesis for zero-shot voice cloning}. The key challenge is to endow a pretrained cloning model with this capability without rebuilding its architecture or relying on explicit style controls. While existing works provide various mechanisms for controlling emotion, style, and voice characteristics~\cite{qwentts, indextts2, moss-vd, hu2026voicesculptor,voxcpm2,fireredtts3}, our focus is complementary: we investigate whether post-training can transform reference-conditioned imitation into text-conditioned expressive generation. More importantly, we study how different post-training paradigms contribute to this transformation and whether they can be combined progressively.

Our study follows three questions. \textbf{First, how should the model be cold-started for this new objective?} We investigate preference alignment with Direct Preference Optimization~\cite{mpo,fpo} (DPO), together with supervised fine-tuning and speaker-diversity ablations, to understand how expressive behavior can be introduced without sacrificing linguistic capability. \textbf{Second, how can the capability ceiling be further improved?} We apply multi-objective Group Sequence Policy Optimization~\cite{gspo} (GSPO) to jointly optimize text-conditioned prosody, acoustic quality, intelligibility, and speaker consistency. \textbf{Third, how can the model become less dependent on the expressiveness of the reference itself?} We use context-conditioned On-Policy Self-Distillation~\cite{opsd} (OPSD), where the model learns under weak-reference conditions from predictive distributions obtained under stronger references, enabling further improvement without requiring stronger references at inference time.

Our post-training stages systematically address SFT degradation, enhance acoustic quality, and enable weak-reference synthesis. Consequently, the final model boosts Emotion from 18.5 to 81.2 and Paralinguistics from 38.4 to 73.9 on Emergent TTS-Eval~\cite{emergent}. Human ABX tests confirm reduced reliance on reference expressiveness, validating post-training as a practical pathway for text-adaptive, expressive zero-shot TTS.


Our contributions are:




\begin{enumerate}
\item We formulate \textbf{text-adaptive prosody synthesis for zero-shot voice cloning} as a distinct research problem, highlighting the gap between faithfully cloning reference speech and generating prosody appropriate to the target text.

\item We develop a \textbf{three-stage post-training framework} that progressively improves text-adaptive speech generation: Direct Preference Optimization (DPO) provides a cold start for expressive alignment, multi-objective Group Sequence Policy Optimization (GSPO) further expands the capability ceiling, and context-conditioned On-Policy Self-Distillation (OPSD) reduces dependence on reference expressiveness.

\item We provide a \textbf{systematic empirical study of post-training for text-adaptive voice cloning}, revealing the effects of speaker diversity, reward composition, and distillation strategy, and demonstrating consistent gains in expressive generation and robustness to weak references on Qwen3-TTS.

\end{enumerate}

\begin{figure*}[th]
    \centering
    \includegraphics[width=1.0\linewidth]{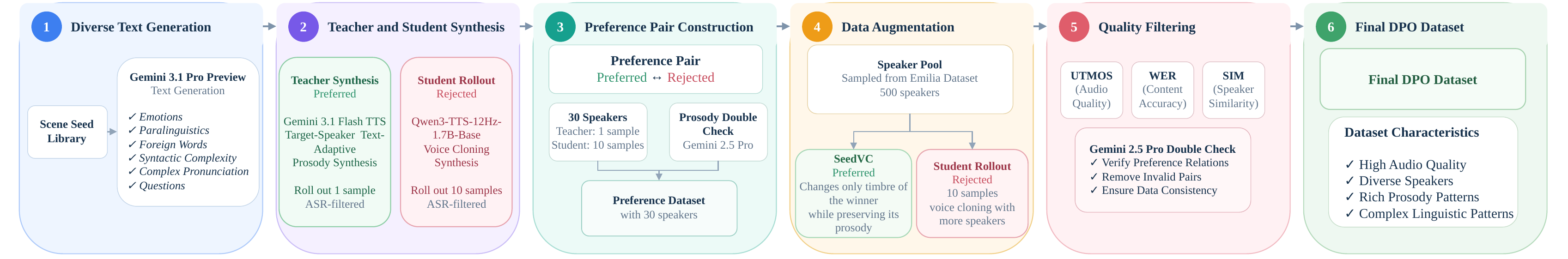}
    \caption{Automated construction of preference data for text-adaptive prosody synthesis.}
    \label{fig:data_pipeline}
\end{figure*}
\section{METHODS}


\begin{table*}[t]
\caption{Main results. Emergent TTS benchmark dimensions are evaluated across In-Set (In, 5 seen speakers), Out-Set (Out, 5 unseen speakers), and Overall (All) average. Objective metrics (WER, SIM, UTMOS) are evaluated on the Seed-TTS-Eval benchmark. Subjective MOS metrics (Naturalness, Prosody, Expressiveness, Speaker Similarity) are reported with 95\% confidence intervals. Stage setup: SFT/DPO-Exp1 (150k audio, 30 spks); DPO-Exp2/GSPO (20k, 500 spks); OPSD (2k, 500 spks). Green highlights denote final results for each stage.}
\label{tab:main_results_speaker_split}
\centering

\scriptsize
\setlength{\tabcolsep}{1.8pt}
\renewcommand{\arraystretch}{1.15}

\begin{tabular}{lcccccccccccccccc}
\toprule
 & \multicolumn{3}{c}{\textbf{Emergent Emotion ($\uparrow$)}} 
 & \multicolumn{3}{c}{\textbf{Emergent Paraling. ($\uparrow$)}} 
 & \multicolumn{6}{c}{\textbf{Objective Metrics}} 
 & \multicolumn{4}{c}{\textbf{Subjective MOS ($\uparrow$)}} \\
\cmidrule(lr){2-4} \cmidrule(lr){5-7} \cmidrule(lr){8-13} \cmidrule(lr){14-17}

\textbf{Model / Stage} 
 & \textbf{In} & \textbf{Out} & \textbf{All} 
 & \textbf{In} & \textbf{Out} & \textbf{All} 
 & \multicolumn{3}{c}{\textbf{EN}} & \multicolumn{3}{c}{\textbf{ZH}} 
 & \textbf{NMOS} & \textbf{PMOS} & \textbf{EMOS} & \textbf{SMOS} \\
\cmidrule(lr){8-10} \cmidrule(lr){11-13}

 & & & & & & & \textbf{WER$\downarrow$} & \textbf{SIM$\uparrow$} & \textbf{UTMOS$\uparrow$} & \textbf{WER$\downarrow$} & \textbf{SIM$\uparrow$} & \textbf{UTMOS$\uparrow$} & & & & \\
\midrule

Base Model (Qwen3-TTS) & 32.1 & 4.9 & 18.5 & 46.0 & 30.7 & 38.4 & \textbf{1.23} & \textbf{0.72} & 3.35 & 0.97 & \textbf{0.77} & 3.17 & \textbf{4.15}{\tiny$\pm$.06} & 3.45{\tiny$\pm$.09} & 3.40{\tiny$\pm$.05} & \textbf{4.00}{\tiny$\pm$.07} \\
\midrule

+ Stage 1 Exp 0: SFT & 82.7 & 36.3 & 59.5 & 67.1 & 40.7 & 53.9 & 1.66 & 0.60 & \textbf{3.47} & \textbf{0.88} & 0.71 & \textbf{3.31} & 4.05{\tiny$\pm$.11} & 3.70{\tiny$\pm$.04} & 3.60{\tiny$\pm$.08} & 3.90{\tiny$\pm$.10} \\
+ Stage 1 Exp 1: DPO & 51.0 & 10.1 & 30.6 & 54.6 & 32.0 & 43.3 & 2.17 & 0.66 & 3.27 & 0.92 & 0.73 & 3.09 & 3.95{\tiny$\pm$.09} & 3.55{\tiny$\pm$.06} & 3.60{\tiny$\pm$.09} & 3.95{\tiny$\pm$.09} \\

\rowcolor{green!10}+ Stage 1 Exp 2: DPO & 83.3 & 50.3 & 66.8 & 71.0 & 44.1 & 57.6 & 1.65 & 0.55 & 3.09 & 0.98 & 0.70 & 3.09 & 3.85{\tiny$\pm$.13} & 3.80{\tiny$\pm$.11} & 3.70{\tiny$\pm$.14} & 3.80{\tiny$\pm$.13} \\

\rowcolor{green!10}+ Stage 2: GSPO & 84.5 & 54.6 & 69.6 & 72.4 & 60.2 & 66.3 & 1.51 & 0.62 & 3.32 & 0.97 & 0.69 & 3.22 & \textbf{4.15}{\tiny$\pm$.10} & 3.90{\tiny$\pm$.05} & \textbf{3.75}{\tiny$\pm$.06} & 3.95{\tiny$\pm$.08} \\

\rowcolor{green!10}+ Stage 3: OPSD & \textbf{90.0} & \textbf{72.4} & \textbf{81.2} & \textbf{80.3} & \textbf{67.4} & \textbf{73.9} & 1.55 & 0.59 & 3.36 & \textbf{0.88} & 0.67 & 3.26 & 4.00{\tiny$\pm$.07} & \textbf{3.95}{\tiny$\pm$.08} & \textbf{3.75}{\tiny$\pm$.10} & 3.80{\tiny$\pm$.06} \\

\bottomrule
\end{tabular}
\end{table*}

We first define the synthesis task, then describe the automated preference data pipeline (Fig.~\ref{fig:data_pipeline}) and the three-stage post-training framework.

\subsection{Task Formulation}

Given a target text $x$ and reference speech $a_{\text{ref}}$, a zero-shot TTS model generates speech according to $y \sim \pi_\theta(y \mid x, a_{\text{ref}})$. The reference conveys both speaker identity $s$ and prosodic characteristics $p_{\text{ref}}$. Although the reference provides cues for reproducing the target voice, its prosody may be unsuitable for the target text. For example, a reference with limited intonational variation may lead to similarly flat speech even when the text calls for an expressive delivery.

We study \emph{timbre-preserving text-adaptive prosody synthesis}: the goal is to retain the speaker identity $s$ while generating prosody $p_y$ appropriate to the semantic and emotional context of $x$. In particular, the model should produce suitable prosody even when expressive cues in $a_{\text{ref}}$ are limited. For brevity, we denote the full conditioning context by $c=(x,a_{\text{ref}})$.

\subsection{Automated Preference Data Construction}
\label{sec:data_pipeline}

We construct preference pairs using teacher-generated speech, student rollouts, voice conversion, and automated quality assessment. Figure~\ref{fig:data_pipeline} summarizes the construction process and the resulting dataset.

\noindent\textbf{Step 1: Text generation.}
Starting from a curated library of scenario seeds, we prompt Gemini 3.1 Pro Preview\footnote{\url{https://ai.google.dev/gemini-api/docs/models/gemini-3.1-pro-preview?hl=zh-cn}} to generate texts covering six dimensions: emotions, paralinguistics, foreign words, syntactic complexity, complex pronunciation, and questions. These texts provide varied linguistic contexts and expressive requirements for synthesis.

\noindent\textbf{Step 2: Teacher and student synthesis.}
For each text--speaker pair drawn from a pool of 30 seed speakers, we obtain one teacher-generated candidate from Gemini 3.1 Flash TTS\footnote{\url{https://ai.google.dev/gemini-api/docs/generate-content/speech-generation?hl=zh-cn}} and ten student rollouts from Qwen3-TTS-12Hz-1.7B-Base~\cite{qwentts}. The student is conditioned on the reference speech associated with the target speaker. Both sets of candidates undergo ASR-based filtering for transcription accuracy.

\noindent\textbf{Step 3: Preference verification.}
We form candidate pairs by assigning teacher and student outputs as the preferred sample $y_w$ and dispreferred sample $y_l$, respectively. Gemini 2.5 Pro\footnote{\url{https://ai.google.dev/gemini-api/docs/models/gemini-2.5-pro?hl=zh-cn}} then checks whether each proposed preference is supported by the prosodic appropriateness and expressiveness of the two samples for the target text. This verification determines whether the proposed preference is valid before the pair is included in the dataset.

\noindent\textbf{Step 4: Speaker augmentation.}
To broaden speaker coverage, we use SeedVC~\cite{seedvc} to convert preferred samples to target voices from a pool of 500 speakers sampled from Emilia~\cite{emilia}. Voice conversion provides additional realizations of the teacher-generated utterances in different voices, with the aim of retaining their content and prosody. We also generate student rollouts using the same target texts and the corresponding target-speaker references to form candidate preference pairs for the expanded speaker pool.

\noindent\textbf{Step 5: Quality filtering.}
We filter candidate pairs using UTMOS~\cite{utmos} for acoustic quality, word error rate (WER) for transcription accuracy, and speaker similarity (SIM) for consistency with the target reference. A final assessment by Gemini 2.5 Pro verifies the preference relations after augmentation. The retained pairs form the final dataset, $\mathcal{D}=\{(c,y_w,y_l)\}$, corresponding to Step 6 in Fig.~\ref{fig:data_pipeline}.

\subsection{Three-Stage Post-Training Framework}
\label{sec:post_training_framework}

Starting from the pretrained Qwen3-TTS model, we apply three stages: DPO on the constructed preference pairs, GSPO with multiple reward objectives, and OPSD across reference conditions with different levels of expressiveness.

\noindent\textbf{Stage 1: Preference alignment with DPO.}
We first optimize the model on $\mathcal{D}$ using DPO. Let $\pi_{\text{ref}}$ denote the frozen pretrained model and define the log-probability ratio as $\ell_\theta(y\mid c)=\log[\pi_\theta(y\mid c)/\pi_{\text{ref}}(y\mid c)]$. The objective is
\begin{equation}
\begin{aligned}
\mathcal{L}_{\text{DPO}}(\theta)
&= -\mathbb{E}_{(c,y_w,y_l)\sim\mathcal{D}}\Bigl[ \\
&\quad \log\sigma\!\left(\beta\left[\ell_\theta(y_w\mid c)
-\ell_\theta(y_l\mid c)\right]\right)\Bigr],
\end{aligned}
\end{equation}
where $\sigma$ is the sigmoid function and $\beta>0$ controls the strength of regularization relative to $\pi_{\text{ref}}$ in the underlying preference optimization formulation. This objective increases the relative likelihood of preferred speech under the same text and reference conditions. We compare DPO with SFT and their combination, and examine the effect of speaker diversity, in the experiments.

\noindent\textbf{Stage 2: Multi-objective alignment with GSPO.}
We next apply GSPO~\cite{gspo} to jointly optimize prosodic expressiveness, acoustic quality, intelligibility, and speaker similarity. For each context $c$, the current policy generates a group of $N$ candidate sequences. Each candidate receives a composite reward:
\begin{equation}
\begin{aligned}
R_{\text{total}}(y;c)
&= w_{\text{prosody}}R_{\text{prosody}}
+ w_{\text{UTMOS}}R_{\text{UTMOS}} \\
&\quad + w_{\text{WER}}R_{\text{WER}}
+ w_{\text{SIM}}R_{\text{SIM}},
\end{aligned}
\end{equation}
where $R_{\text{prosody}}$ is the Gemini 2.5 Pro assessment of text-conditioned prosodic expressiveness, $R_{\text{UTMOS}}$ measures acoustic quality, $R_{\text{WER}}$ is a WER-based intelligibility reward, and $R_{\text{SIM}}$ measures speaker similarity to $a_{\text{ref}}$. The weights control the contributions of the four objectives. GSPO uses group-normalized rewards to estimate relative advantages and performs sequence-level policy updates. Joint optimization is intended to balance prosodic gains with acoustic quality, content accuracy, and speaker consistency.


\noindent\textbf{Stage 3: Context-conditioned On-Policy Self-Distillation.}
To reduce sensitivity to reference expressiveness, we apply OPSD ~\cite{opsd}
with a weak-reference context $c_w=(x,a_{\mathrm{ref}}^{\mathrm{weak}})$
and a strong-reference context $c_s=(x,a_{\mathrm{ref}}^{\mathrm{strong}})$.
At decoding step $t$, let
$p_t=\pi_\theta(\cdot\mid y_{<t},c_w)$ and
$q_t=\pi_{\mathrm{teacher}}(\cdot\mid y_{<t},c_s)$. The student is
trained to match the teacher using full-vocabulary JSD:
\begin{equation}
\begin{aligned}
\mathcal{L}_{\mathrm{OPSD}}
&=
\frac{1}{2T}\sum_{t=1}^{T}
\Bigl[
D_{\mathrm{KL}}(p_t\|m_t)
+
D_{\mathrm{KL}}(q_t\|m_t)
\Bigr],\\
m_t &= \frac{p_t+q_t}{2},
\end{aligned}
\label{eq:opsd}
\end{equation}
where the teacher distribution is detached. We also compare a
sampled-token objective, where $y_t\sim p_t$ and
\begin{equation}
\mathcal{L}_{\mathrm{ST}}
=
-\frac{1}{T}\sum_{t=1}^{T}
\operatorname{sg}\!\left[
\log q_t(y_t)-\log p_t(y_t)
\right]\log p_t(y_t).
\label{eq:sampled_token}
\end{equation}
The sampled-token variant avoids full-vocabulary computation.

\begin{table}[t]
\caption{Ablation on Stage 1 data strategy and speaker diversity scaling evaluated across 6 Emergent TTS dimensions on an In-Domain speaker. Dimensions: \textbf{Emo.} (Emotions), \textbf{For.} (Foreign Words), \textbf{Par.} (Paralinguistics), \textbf{Pro.} (Complex Pronunciation), \textbf{Que.} (Questions), \textbf{Syn.} (Syntactic Complexity).}
\label{tab:stage1_ablation_single_col}
\centering
\renewcommand{\arraystretch}{1.15}
\footnotesize
\setlength{\tabcolsep}{2pt}
\begin{tabularx}{\linewidth}{l l *{6}{>{\centering\arraybackslash}X}}
\toprule
\textbf{ID} & \textbf{Setting / Strategy} & \multicolumn{6}{c}{\textbf{Emergent TTS Benchmark ($\uparrow$)}} \\
\cmidrule(lr){3-8}
 & & \textbf{Emo.} & \textbf{For.} & \textbf{Par.} & \textbf{Pro.} & \textbf{Que.} & \textbf{Syn.} \\
\midrule
Top & Gemini 3.1 Flash TTS & 88.2 & 50.9	& \textbf{90.4}	& \textbf{71.2} &	73.4	& \textbf{64.6} \\
\midrule

\multicolumn{8}{l}{\textit{Part A: Gemini Distilled Data (150k, 30 Speakers)}} \\

Exp 0 & Base Model                   & 65.2 & 33.0 & 70.0 & 65.3 & 72.7 & 56.8 \\

Exp 1 & SFT Only                     & 82.3 & 43.7 & 70.3 & 53.3 & 73.3 & 61.1 \\
Exp 2 & DPO Only                     & 81.4 & 52.7 & 78.4 & 69.1 & 71.4 & 55.6 \\

Exp 3 & DPO + SFT ($\lambda{=}0.1$)   & 82.6 & 50.0 & 74.1 & 67.4 & 75.7 & 58.5 \\
Exp 4 & DPO + SFT ($\lambda{=}0.5$)   & 81.9 & 39.6 & 74.6 & 66.9 & 70.4 & 61.9 \\

\midrule

\multicolumn{8}{l}{\textit{Part B: SeedVC Preference Data (20k, Diversity Scaling)}} \\
Exp 5 & DPO (100 Spk)             & 86.9 & 49.6 & 80.2 & 62.0 & \textbf{76.8}& 57.1 \\
Exp 6 & DPO (500 Spk) & \textbf{89.8} & \textbf{59.3} & 82.0 & 64.3 & 73.9 & 60.2   \\
\bottomrule
\end{tabularx}
\end{table}

\begin{table}[t]
\caption{Ablation studies on Stage 2 GSPO reward design and Stage 3 OPSD distillation strategies on an In-Domain speaker. \textbf{Part A} investigates GSPO hyper-parameters (Number of rollouts $N$, Batch size $B$, Default: $N{=}4, B{=}64$) and reward components ($\omega_{\text{Gemini}}{=}1.0$, $\omega_{\text{SIM}}{=}1.0$, others $0.1$). 
\textbf{Part B} evaluates OPSD, comparing Full-Vocabulary Logit Distillation against Sampled-Token Distillation. Objective metrics are evaluated on the  standard Seed-TTS-Eval EN benchmark.}
\label{tab:combined_ablation}
\centering
\renewcommand{\arraystretch}{1.15}
\resizebox{\linewidth}{!}{
\begin{tabular}{lccccc}
\toprule
\textbf{Ablation Setting} & \multicolumn{2}{c}{\textbf{Emergent ($\uparrow$)}} & \multicolumn{3}{c}{\textbf{Objective Metrics}} \\
\cmidrule(lr){2-3} \cmidrule(lr){4-6}
 & \textbf{Emo.} & \textbf{Para.} & \textbf{WER $\downarrow$} & \textbf{SIM $\uparrow$} & \textbf{UTMOS $\uparrow$} \\
\midrule
\multicolumn{6}{l}{\textit{\textbf{Part A: Stage 2 GSPO Ablation}}} \\
\rowcolor{green!10} Full GSPO (Default: $N{=}4, B{=}64$) & \textbf{92.3} & 82.1  & \textbf{1.51} & 0.62 & 3.32 \\

\quad \textit{1) Rollout/Batch Sensitivity:} & & & & & \\
\quad $N{=}64, B{=}4$ & 90.5 & 80.5 &  1.74 & 0.62 & 3.19 \\
\quad \textit{2) Reward Component Analysis:} & & & & & \\
\quad w/o Gemini Prosody ($\omega{=}0$) & 62.9 & 65.4 & 1.52 & \textbf{0.66} & \textbf{3.42} \\
\quad w/o UTMOS ($\omega{=}0$) & 80.0 & 71.1 & 1.56 & 0.62 & 2.74 \\

\midrule
\multicolumn{6}{l}{\textit{\textbf{Part B: Stage 3 OPSD Ablation}}} \\
\quad Sampled-Token & 90.0 & 81.8 & 1.53 & 0.57 & 3.38 \\
\rowcolor{green!10} \quad Full-Vocab, JSD ($\beta{=}0.5$) & \textbf{91.3} & \textbf{82.3} & 1.55 & \textbf{0.59} & 3.36 \\
\bottomrule
\end{tabular}
}
\end{table}

\section{EXPERIMENTS}

\subsection{Implementation Details}
We adopt the following training schedule for the three post-training stages.

\noindent\textbf{Stage 1: SFT and DPO.}
All Stage 1 variants, including SFT, DPO, and their combination, are trained for three epochs.

\noindent\textbf{Stage 2: Multi-objective GSPO.}
Starting from the best-performing Stage 1 checkpoint, we perform one epoch of GSPO to jointly optimize prosodic expressiveness, acoustic quality, intelligibility, and speaker similarity.

\noindent\textbf{Stage 3: Context-conditioned OPSD.}
We initialize the model from the Stage 2 checkpoint and train it for one additional epoch using OPSD, with the aim of improving prosodic expressiveness under weak-reference conditions.





\subsection{Main Results}

Table~\ref{tab:main_results_speaker_split} summarizes results on Emergent TTS-Eval, Seed-TTS-Eval~\cite{seed-tts}, and subjective listening tests. For Emergent TTS-Eval, we report performance on seen speakers (In-Set), unseen speakers (Out-Set), and their overall average.

\noindent\textbf{Stage 1: Cold-start alignment.}
Under the 150k, 30-speaker configuration, SFT outperforms DPO in overall Emotion (59.5 vs.\ 30.6) and Paralinguistics (53.9 vs.\ 43.3), although both exhibit substantial gaps between seen and unseen speakers. With 20k preference pairs covering 500 speakers, DPO achieves higher overall scores of 66.8 and 57.6, respectively, and improves Out-Set Emotion to 50.3. These results highlight the effectiveness of the speaker-diverse data configuration. However, because dataset size and construction also differ, this comparison alone does not isolate the effect of speaker diversity.

\noindent\textbf{Stage 2: Multi-objective GSPO.}
Starting from the 500-speaker DPO checkpoint, GSPO improves overall Emotion from 66.8 to 69.6 and Paralinguistics from 57.6 to 66.3. These gains are accompanied by improvements in English UTMOS from 3.09 to 3.32. Thus, GSPO improves both measured prosodic performance and acoustic quality relative to the DPO checkpoint.

\begin{figure}[th]
    \centering
    \includegraphics[width=0.95\linewidth]{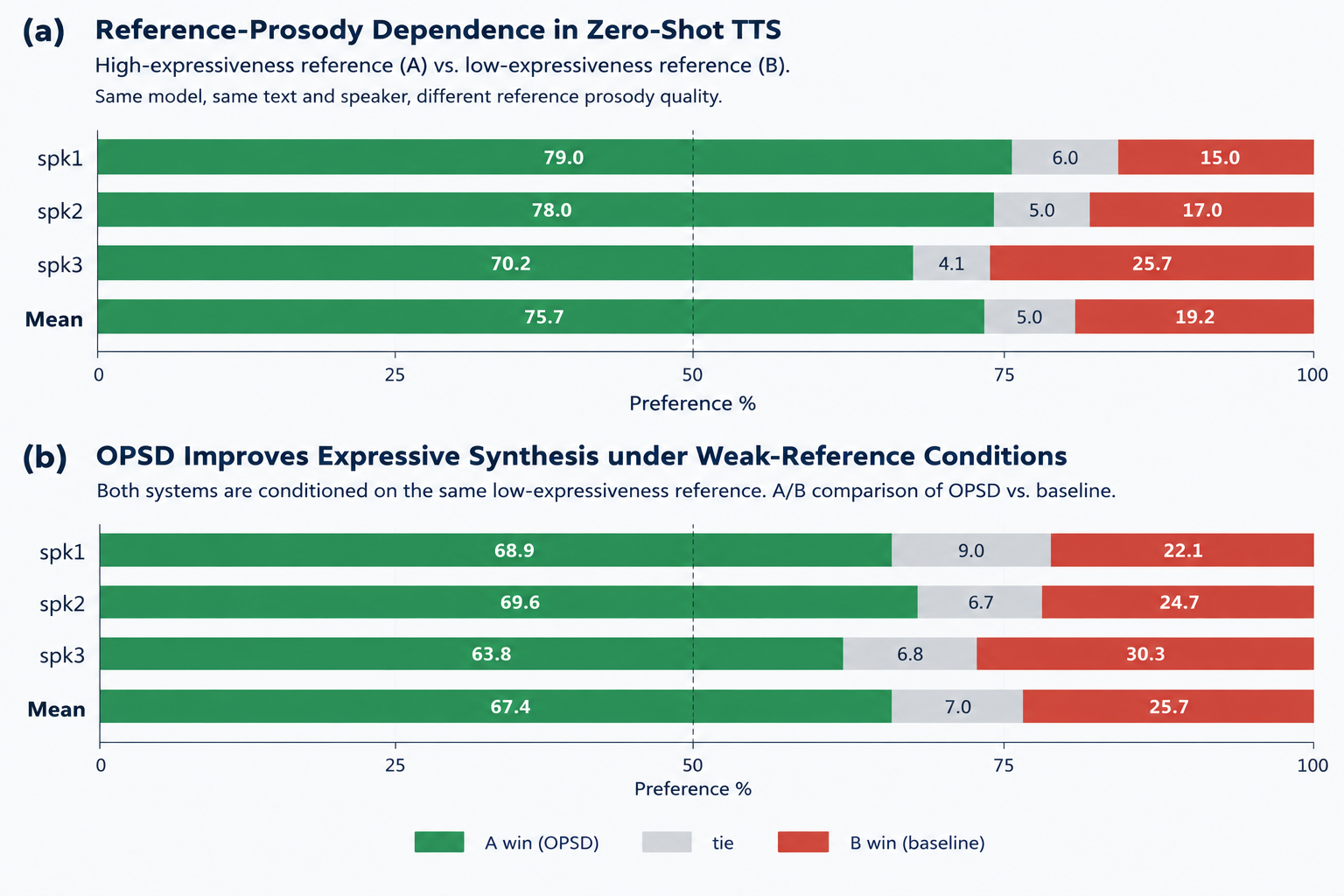}
    \caption{Human ABX preference results for reference-prosody dependence and OPSD robustness, evaluated on 3360 ABX pairs by 20 listeners. (a) High- vs. low-expressiveness references. (b) OPSD vs. baseline under low-expressiveness references.}
    \label{fig:opsd}
\end{figure}

\noindent\textbf{Stage 3: Context-conditioned OPSD.}
OPSD further improves overall Emotion and Paralinguistics to 81.2 and 73.9, respectively, the highest scores among the evaluated configurations. Out-Set Emotion increases from 54.6 to 72.4, indicating improved performance on unseen speakers. Subjective PMOS show smaller numerical gains, increasing from 3.90 to 3.95 respectively. These improvements are accompanied by a decrease in NMOS from 4.15 to 4.00 and lower speaker similarity.
To directly assess reference-prosody dependence, we conduct a human ABX preference study (Fig.~\ref{fig:opsd}). Under matched model, text, and speaker conditions, high-expressiveness references are preferred over low-expressiveness references by 75.7\% vs.~19.2\%, confirming the baseline's sensitivity to reference prosodic quality. When both systems are conditioned on the same low-expressiveness reference, OPSD is preferred over the baseline by 67.4\% vs.~25.7\% on average, with consistent gains across all three speakers. These results directly support the robustness of OPSD under weak-reference conditions.

\noindent\textbf{Speaker similarity.}
Relative to the base model, the final model's SIM decreases from 0.72 to 0.59 in English and from 0.77 to 0.67 in Chinese. Subjective SMOS also decreases from 4.00 to 3.80. Although changes in prosody may influence automatic similarity measurements, the accompanying decline in SMOS suggests that the reduction cannot be attributed solely to metric sensitivity. These results reveal a remaining trade-off between expressive prosody and speaker similarity, making stronger identity preservation an important direction for further improvement.

\noindent\textbf{Agreement with human preferences.}
In the A/B evaluation, preferences predicted by Gemini 2.5 Pro agree with human judgments on 80.1\% of the evaluated pairs. This agreement provides supporting evidence that the automated evaluator captures aspects of human-perceived prosodic quality within the evaluated setting.

\subsection{Ablation Studies}

\noindent\textbf{Stage 1: Data strategy and speaker diversity.} Table~\ref{tab:stage1_ablation_single_col} (Part A) compares training objectives on Gemini-distilled data. While SFT boosts Emotion ($65.2\rightarrow82.3$) at the cost of Complex Pronunciation ($65.3\rightarrow53.3$), DPO avoids this degradation (81.4 Emotion, 69.1 Pronunciation). Combining DPO with SFT ($\lambda=0.1$) balances both (82.6 Emotion, 67.4 Pronunciation). In Part B (20k SeedVC pairs), expanding speakers from 100 to 500 improves 5 of 6 dimensions, e.g., Emotion ($86.9\rightarrow89.8$), despite Questions dropping ($76.8\rightarrow73.9$), confirming that broader speaker diversity enhances preference optimization under a fixed pair budget.

\noindent\textbf{Stage 2: GSPO configurations and rewards.}
Table~\ref{tab:combined_ablation}, Part A, examines two rollout and batch configurations. Compared with $N=64, B=4$, the default configuration ($N=4, B=64$) achieves higher Emotion (92.3 vs.\ 90.5) and Paralinguistics (82.1 vs.\ 80.5), together with lower WER (1.51 vs.\ 1.74). Reward ablations further show that removing the Gemini prosody reward reduces Emotion from 92.3 to 62.9 and Paralinguistics from 82.1 to 65.4. Removing the UTMOS reward reduces UTMOS from 3.32 to 2.74. These results support the contributions of the prosody and acoustic-quality rewards to the combined objective.

\noindent\textbf{Stage 3: OPSD distillation strategy.}
Table~\ref{tab:combined_ablation}, Part B, compares sampled-token distillation with full-vocabulary JSD at $\lambda=0.5$. Full-vocabulary JSD achieves higher Emotion (91.3 vs.\ 90.0), Paralinguistics (82.3 vs.\ 81.8), and SIM (0.57 vs.\ 0.59). These gains are accompanied by a slightly higher WER (1.55 vs.\ 1.53) and lower UTMOS (3.36 vs.\ 3.38). The comparison therefore favors full-vocabulary JSD for measured expressiveness and speaker similarity, while highlighting a trade-off with intelligibility and acoustic quality.





\section{CONCLUSION}

We presented a three-stage post-training framework for text-adaptive prosody synthesis, with the goal of preserving speaker identity while reducing dependence on reference prosody. The framework combines DPO, multi-objective GSPO, and context-conditioned OPSD. Our ablations show that DPO avoids the complex-pronunciation degradation observed with SFT in the evaluated setting, while broader speaker coverage benefits preference alignment. GSPO improves both prosodic performance and acoustic quality relative to the DPO checkpoint, and OPSD further increases prosodic scores. These improvements are accompanied by reduced speaker similarity, revealing a remaining trade-off between expressiveness and identity preservation. Future work should strengthen speaker consistency and systematically evaluate robustness across reference conditions.

\clearpage
\balance
\bibliographystyle{IEEEbib}
\bibliography{refs}

\end{document}